\documentstyle[twocolumn,aps,prc,psfig]{revtex}
\begin{document}
\title{
Possibility to deduce the emission time sequence \\
of neutrons and protons from the neutron-proton 
correlation function?
}
\author{   
  R.~Ghetti$^{a,1}$,
  J.~Helgesson$^{b}$,
  N.~Colonna$^{c}$,
  B.~Jakobsson$^{a}$,
  A.~Anzalone$^{d}$,  
  V.~Bellini$^{d,e}$,   
  L.~Carl\`{e}n$^{a}$,
  S.~Cavallaro$^{d,e}$, 
  L.~Celano$^{c}$,
  E.~De Filippo$^{e,f}$, 
  G.~D'Erasmo$^{c}$, 
  D.~Di Santo$^{c}$,
  E.M.~Fiore$^{c}$, 
  A.~Fokin$^{a}$, 
  M.~Geraci$^{f}$, 
  F.~Giustolisi$^{e,f}$,
  A.~Kuznetsov$^{g}$, 
  G.~Lanzan\`{o}$^{e,f}$,
  D.~Mahboub$^{d}$, 
  S.~Marrone$^{c}$, 
  J.~M{\aa}rtensson$^{a}$,
  M.~Palomba$^{c}$,  
  A.~Pantaleo$^{c}$,
  V.~Paticchio$^{c}$,
  G.~Riera$^{d}$,
  M.L.~Sperduto$^{d,e}$, 
  C.~Sutera$^{e,f}$, 
  G.~Tagliente$^{c}$, 
  M.~Urrata$^{d}$\\
{\small \it 
$^a$Department of Physics, Lund University, 
Box 118, SE-221 00 Lund, Sweden \\
$^b$Malm\"{o} University, 
School of Technology and Society, 
SE-205 06 Malm\"{o}, Sweden \\
$^c$INFN and Dipartimento di Fisica, 
Via Amendola 173, I-70126 Bari, Italy\\
$^d$Laboratori Nazionali del Sud (INFN), Via S. Sofia 44, 
I-95123 Catania, Italy\\
$^e$Dipartimento di Fisica, Universit\`{a} di Catania, 
Corso Italia 57, I-95129 Catania, Italy\\
$^f$INFN, Sezione di Catania,
Corso Italia 57, I-95129 Catania, Italy\\
$^g$Khlopin Radium Institut, 
Shvernik Avenue\ 28, 194021 St.\ Petersburg, Russia
}
}
\date{\today}
\maketitle
\begin{abstract}
Experimental information has been derived from the neutron-proton 
correlation function in order to deduce the time sequence 
of neutrons and protons emitted at 45$^{\rm o}$ in 
the E/A = 45 MeV $^{58}$Ni + $^{27}$Al reaction.
\end{abstract}

\vspace{0.2cm}
\noindent
PACS number(s): 25.70.Pq, 29.30.Hs 

\vspace{0.2cm}
{\small 
\noindent
$^1$Corresponding Author: Roberta Ghetti, 
Department of Physics, Lund University, 
Box 118, SE-221 00 Lund, Sweden
Tel.\ +46-46-2227388, Fax +46-46-2224015}\\

\normalsize
Two-nucleon correlation functions are normally utilized 
to extract information on the size of the emitting source 
and on the time duration of the emission \cite{Rev}. 
Furthermore, when correlations 
of non-identical particles are considered, additional 
model-independent information on the emission chronology of the 
particles can be obtained. 

A technique to probe the emission sequence and time delay of ejectiles 
in nuclear reactions was first suggested for charged particle 
pairs \cite{NIM94}, based on the idea that mutual Coulomb 
repulsion would be experienced by pairs of 
charged particles emitted with a short time delay. Comparison of 
the velocity difference spectra with 
trajectory calculations would thus give a measure of the average 
particle emission sequence \cite{Gelderloos}.
The technique was extended to any kind of 
interacting, non-identical particle 
pairs in the theoretical study of Ref.\ \cite{Lednicky}. 
There it was demonstrated that the sensitivity 
of the correlation function to the asymmetry of the 
distribution of the relative space-time coordinates of the particle 
emission points can be used to determine the differences in the 
mean emission times.
This effect has been discussed for different particle pairs,  
for instance $pd$, $np$ \cite{Lednicky}, $\pi p$ \cite{Volo97}, 
$K^+K^-$ \cite{Ardo99}. 
The technique of Ref.\ \cite{Lednicky} has been applied in 
recent experimental studies where the time sequence 
for $p$ and $d$ emitted in the E/A = 50 MeV Xe + Sn reaction \cite{Gourio} 
and for $p, d, t, ^3He$ and $\alpha$ particles 
emitted in central E/A = 400 MeV Ru + Zr collisions \cite{Kotte} 
has been deduced. Typical time delays of a few fm/c have been 
found in \cite{Kotte}.  

In this paper we report on the first experimental evidence that 
the emission chronology of neutrons and protons can be deduced 
from the neutron-proton ($np$) correlation function. These data come 
from the  E/A = 45 MeV $^{58}$Ni + $^{27}$Al 
reaction \cite{NPA99,NPA00,Brief}. 
After a short discussion about the possibility to obtain 
chronology information, clarified with the help of 
theoretical calculations of the effects, 
indications on the emission sequence are derived 
from the experimental ungated $np$ correlation function 
as well as from the correlation function gated on the parallel 
velocity and on the total momentum of the nucleon pair.

The possibility to access information from the $np$ 
correlation function on the particle emission time sequence 
is connected to the different strength of the nuclear final 
state interaction experienced by different pairs. 
Namely, from a merely classical viewpoint, 
if there are pairs for which the average distance 
between the two particles (when the two particles 
start to interact) is smaller, these pairs experience a 
stronger interaction and exhibit an enhanced correlation 
(or anticorrelation) strength, as compared to those pairs 
for which the average particle distance is larger. 
Now, if there is an average time difference in the 
emission times between the two particles, there will also be 
a difference in the average distance for 
the two ``classes'' $E_n > E_p$ and $E_n < E_p$.

In order to study the $n-p$ emission chronology, 
one can calculate the  $np$ correlation function with different 
source models and parameter sets. 
Due to the strong final state interaction between $n$ and $p$, 
the $np$ correlation function shows a strong 
correlation for small values of the relative momentum $q$ and, 
for some parameter sets, 
a weak  anticorrelation for intermediate values of $q$. 
This is due to the fact that the correlation 
in the singlet channel (antiparallel spins of $np$) is positive while 
it is negative in the triplet channel (parallel spins of $np$) 
even if the interaction in both channels is attractive. 
The counterintuitive negative correlation in the triplet channel is
due to the deuteron formation. Because of this process, the 
sample of $np$ pairs is depleted and 
one effectively observes the negative correlation \cite{Stanislaw}.

If a time delay in the emission of the two particles is introduced, 
the interaction is enhanced for those pairs for which the 
average distance is smaller. 
This can be easily seen if one compares the correlation 
function $C_n(q)$, gated on pairs $E_n > E_p$, 
with the correlation function $C_p(q)$, gated on pairs $E_n < E_p$. 
If the proton is emitted earlier (later) than the neutron, 
the ratio $C_n/C_p$ will show a peak (dip) in the region of $q$ where 
there is a correlation, a dip (peak) where there is an anticorrelation, 
and will approach unity both for $q \rightarrow$ 0 
(since the energy difference of the two 
emitted particles is negligible) and $q \rightarrow \infty$ 
(since modifications of the two-particle phase space density 
arising from final state interactions are negligible). 
The exact location of the peak and dip in the ratio depends on the 
source model and on the parameter set (as does of course the 
correlation function itself).
 \begin{figure}
 \centerline{\psfig{file=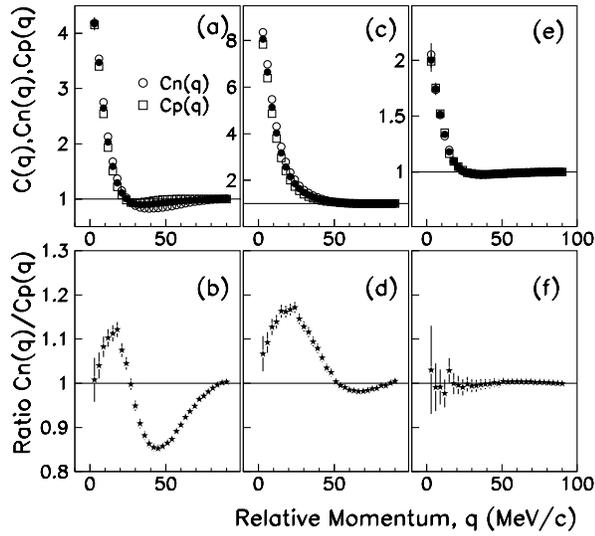,height=8.cm,angle=0}}
 \caption{\rm 
Calculations performed with the source model of Ref.\ [14].
Panels~(a)(b): $np$ correlation functions $C(q)$ (solid dots), 
$C_n(q)$, $C_p(q)$ and the ratio $C_n/C_p$, 
from a source emitting $p$ and $n$ with 
emission time width of $\Delta t$ = 40 fm/c 
and with the neutron emission time delayed 50 fm/c. 
Panels~(c)(d): the same as in panels~(a)(b), 
obtained with emission time width of $\Delta t$ = 80 fm/c. 
Panels~(e)(f): the same as in panels~(a)(b), 
from a ``two-source model'' calculation [10], 
with exponential lifetime of 
$\tau_p \approx$ 400$\pm$200 fm/c for protons, 
$\tau_n \approx$ 600$\pm$200 fm/c for neutrons
($\approx$ 30$\%$ of $p$ and $\approx$ 13$\%$ of $n$ 
come from pre-equilibrium).
 }
 \label{fig:csorgo}
 \end{figure}

This effect has been calculated in Ref.\ \cite{Lednicky} 
for different delays in the emission of neutrons and protons. 
The calculation for the $np$ system assuming 
that protons are on average emitted 100 fm/c earlier than neutrons 
(right panels in Fig.\ 2 of Ref.\ \cite{Lednicky}), 
shows $C_n(q) < 1$ down to $q \approx$ 15 MeV/c and a 
corresponding dip in the ratio $C_n/C_p$. The expected peak 
at small $q$, where the correlation function 
shows a strong correlation, is 
washed out in these calculations, since the statistics is very low 
and the parameter set generates a correlation only 
in the very low $q$-region ($<$ 15 MeV/c), 
where the ratio $C_n/C_p$ has to approach unity \cite{Private}.

To illustrate this point further, we present in Fig.\ \ref{fig:csorgo} 
three model calculations performed with the source model 
of Ref.\ \cite{Helgesson} coupled to the Koonin-Pratt 
formalism \cite{KoonPrat} to include final state interactions.
The first calculation represents the $np$ correlation function 
obtained from the source model of Ref.\ \cite{Helgesson} 
with time width parameter $\Delta t$ = 40 fm/c and with the 
neutron emission time delayed 50 fm/c with respect to the proton
emission time. 
Fig.\ \ref{fig:csorgo}a shows the calculated 
$C(q)$ (solid dots), $C_n(q)$ (open circles) and $C_p(q)$ (open squares). 
The ratio  $C_n/C_p$ is shown in Fig.\ \ref{fig:csorgo}b.
$C(q)$ presents a weak anticorrelation in the intermediate $q$-region 
between 30 and 50 MeV/c and a strong correlation at $q <$ 20 MeV/c.
The correlation function $C_n(q)$ gated on 
pairs $E_n > E_p$ demonstrates an enhanced interaction, since 
the $n-p$ average distance when the particles start to interact 
is shorter for these pairs. The ratio  $C_n/C_p$ is above unity  
where $C(q)$ has a correlation and below unity where  $C(q)$ has an 
anticorrelation
Figs.\ \ref{fig:csorgo}c, \ref{fig:csorgo}d illustrate the 
results of a similar 
calculation with larger width of the time distribution 
$\Delta t$ = 80 fm/c.
One can see that the anticorrelation in $C(q)$ is weaker 
and shifted to larger values of $q$. The ratio $C_n/C_p$ is 
changed accordingly. 
Finally Figs.\ \ref{fig:csorgo}e, \ref{fig:csorgo}f present the results of 
the ``two-source'' calculation of Ref.\ \cite{NPA00}, 
where it is assumed that $\approx$ 30$\%$ of 
the protons and $\approx$ 13$\%$ of the neutrons 
come from pre-equilibrium emission. 
The parameter set used for this calculation yields 
emission time distributions with exponential lifetime 
values of  $\tau_p \approx$ 400$\pm$200 fm/c for protons and 
$\tau_n \approx$ 600$\pm$200 fm/c for neutrons. 
The ratio $C_n/C_q$ remains close to unity for all values of $q$. 

\begin{figure}
 \centerline{\psfig{file=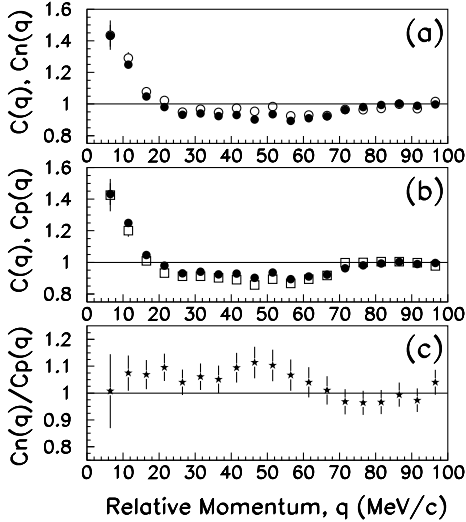,height=8.cm,angle=0}}
 \caption{\rm 
Experimental ungated $np$ correlation function $C(q)$,  
from the E/A = 45 MeV $^{58}$Ni + $^{27}$Al reaction, 
(solid dots in panels~(a)(b)) compared to:  
Panel~(a), open circles: $C_n(q)$, constructed from pairs of type $E_n > E_p$.
Panel~(b), open squares: $C_p(q)$, constructed from pairs of type $E_n < E_p$.
The ratio $C_n/C_p$ is shown in panel~(c). 
 }
 \label{fig:lednicky}
 \end{figure}

The experimental $np$ correlation function from 
the E/A = 45 MeV $^{58}$Ni + $^{27}$Al reaction, obtained with a  
proton detection array placed at 45$^{\rm o}$ and a 
neutron setup of liquid scintillators placed 
at 25$^{\rm o}$, 45$^{\rm o}$ and 90$^{\rm o}$, 
was presented in Ref.\ \cite{NPA00} and is  
shown by the solid dots in Figs.\ \ref{fig:lednicky}a, \ref{fig:lednicky}b. 
Kinetic energy thresholds of 2--50 MeV for neutrons 
and 6--50 MeV for protons are applied to the data. 
The normalization is defined in the region $q$ = 80--120 MeV/c. 
$C(q)$ shows a clear correlation for $q <$ 20 MeV/c 
and a very small anticorrelation for 20$ < q <$ 70 MeV/c. 

In order to study the $n-p$ emission chronology, 
the correlation function $C_n(q)$, gated on pairs $E_n > E_p$, 
and $C_p(q)$,  gated  on pairs $E_n < E_p$,  
are presented in Fig.\ \ref{fig:lednicky} 
together with the ratio $C_n/C_p$. 
$E_n$ and $E_p$ are calculated in the reference frame of a 
source moving with a velocity of 0.16c \cite{Brief}. 
A single normalization constant, calculated from the 
ungated correlation function, is utilized for both 
energy-gated correlation functions $C_n(q)$ and $C_p(q)$.

The data show a small experimental signal.
$C_n(q)$ is slightly enhanced 
with respect to $C(q)$ for all values of $q <$ 60 MeV/c 
(Fig.\ \ref{fig:lednicky}a).
Correspondingly $C_p(q)$ is suppressed in the same $q$-region 
(Fig.\ \ref{fig:lednicky}b). 
The ratio $C_n/C_p$ is above unity for $q <$ 70 MeV/c 
(Fig.\ \ref{fig:lednicky}c).
The signal has proved to be rather stable. 
When testing it against the accuracy of the energy calibrations, 
it turns out that a statistical uncertainty of up to 20--30$\%$ 
does not have any appreciable effects on the  $C_n/C_q$ ratio. 
On the other hand, a systematic error of the same size might 
significantly alter the final results, but we believe that the 
accuracy of our energy calibrations is within 5--10$\%$ \cite{NPA00}. 

As discussed above, 
the enhancement of $C_n(q)$ where $C(q)$ has a correlation 
($q <$ 20 MeV/c) might indicate that for those pairs that 
contribute to the region $q <$ 20 MeV/c protons  
are on average emitted earlier than neutrons.
The enhancement of $C_n(q)$ in the region 20 $< q <$ 60 MeV/c 
might instead indicate the opposite emission time sequence 
as $C(q)$ has an anticorrelation there.
However, one should keep in mind that the inclusive correlation 
function reflects a convolution of different 
sources and origins of emission thus it is not reasonable 
to expect that the information from the inclusive data should 
provide us with a well defined and unique $np$ emission time sequence. 

In order to disentangle the different effects and minimize 
such averaging, various kinds of gates can be 
applied to the inclusive data. 
We shall now present correlation functions with additional 
gates and we shall see that indeed, 
despite the poor statistics, the gated correlation functions 
yield a ratio $C_n/C_p$ in closer accordance with the 
qualitative appearance of Fig.\ \ref{fig:csorgo}. 
 \begin{figure}
 \centerline{\psfig{file=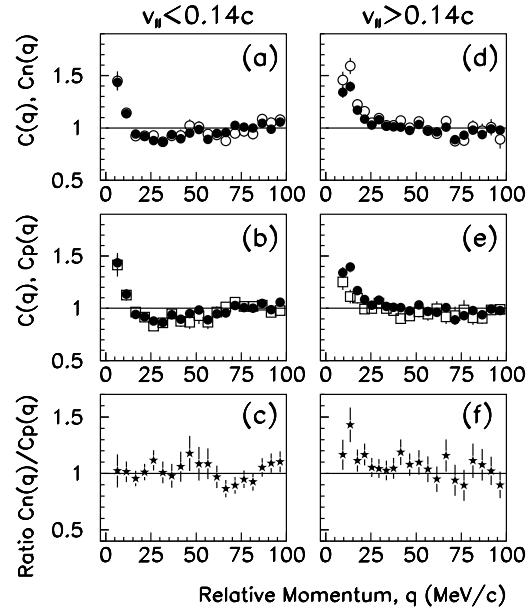,height=9.cm,angle=0}}
 \caption{\rm 
The same as in Fig.\ 2, 
for low-$v_{\parallel}$ ($<$ 0.14c) $np$ pairs 
(panels~(a)(b)(c)) and 
for high-$v_{\parallel}$ ($>$ 0.14c) $np$ pairs 
(panels~(d)(e)(f)).
 }
 \label{fig:lednicky-cuts}
 \end{figure}
Fig.\ \ref{fig:lednicky-cuts} illustrates how the results 
of Fig.\ \ref{fig:lednicky} are modified when cuts on 
the parallel velocity of the particles are applied. 
The effects of these cuts, that are defined on the basis of 
BUU calculations \cite{Sources} and of the available 
statistics, are to enhance (suppress) the quasi-projectile source by 
selecting nucleons with large (small) values of the parallel velocity. 
Figs.\ \ref{fig:lednicky-cuts}a, \ref{fig:lednicky-cuts}b,  
\ref{fig:lednicky-cuts}c correspond to a low-$v_{\parallel}$ gate 
and Figs.\ \ref{fig:lednicky-cuts}d, \ref{fig:lednicky-cuts}e, 
\ref{fig:lednicky-cuts}f to a high-$v_{\parallel}$ gate. 
One can notice that the behavior of the 
correlation functions and of 
$C_n/C_p$ ratio is quite different for the two 
different event selections. 
``Projectile-like'' events (i.e.\ high-$v_{\parallel}$ selection) 
exhibit an enhanced correlation function strength and  
an enhanced $C_n/C_p$ ratio in correspondence to the 
correlation ($q <$ 25 MeV/c). 
``Target-like'' events (i.e.\ low-$v_{\parallel}$ selection) 
exhibit a slight suppression in the correlation function 
strength ($q <$ 20 MeV/c), a weak anticorrelation for 
$q \approx$ 20--50 MeV/c and a $C_n/C_p$ ratio quite close 
to unity in the correlation region. 

\begin{figure}
 \centerline{\psfig{file=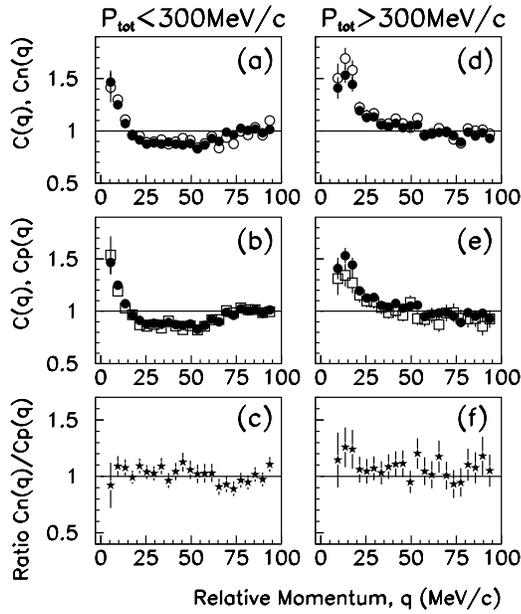,height=9.cm,angle=0}}
 \caption{\rm 
The same as in Fig.\ 2, 
for low-$P_{tot}$ ($<$ 300 MeV/c) $np$ pairs 
(panels~(a)(b)(c)) 
and for high-$P_{tot}$ ($>$ 300 MeV/c) $np$ pairs 
(panels~(d)(e)(f)). 
$P_{tot}$ is calculated in the reference frame of a source 
moving with  $v_{source}$ = 0.16 c. 
 }
 \label{fig:lednicky-cuts-2}
 \end{figure}

Gating on the momentum (or energy) of the particle pair in the 
source system is particularly useful to investigate the time-scale 
of the emission and to study the 
interplay between dynamical and statistical effects in 
particle emission. 
The $np$ correlation function gated on high-$P_{tot}$ pairs 
has already been presented in Ref.\ \cite{NPA00}. 
An enhancement in the correlation function strength was observed 
indicating a smaller space-time extent of the pre-equilibrium source 
of emission \cite{NPA00}.
Fig.\ \ref{fig:lednicky-cuts-2} presents 
total-momentum-gated correlation functions. 
A low-$P_{tot}$ gate is applied to all results shown in 
Figs.\ \ref{fig:lednicky-cuts-2}a, 
\ref{fig:lednicky-cuts-2}b, \ref{fig:lednicky-cuts-2}c 
and a high-$P_{tot}$ gate 
in Figs.\ \ref{fig:lednicky-cuts-2}d, 
\ref{fig:lednicky-cuts-2}e, \ref{fig:lednicky-cuts-2}f. 

The effects of the gates on the correlation functions 
and on the $C_n/C_p$ ratio are similar to those observed 
in Fig.\ \ref{fig:lednicky-cuts}. 
High-total-momentum selected events 
exhibit an enhanced correlation function strength and  
an enhanced $C_n/C_p$ ratio in correspondence to the 
correlation ($q <$ 25 MeV/c). 
Low-total-momentum selected events 
exhibit a slight suppression in the correlation function 
strength ($q <$ 20 MeV/c), a weak anticorrelation in the 
intermediate $q$-region and a $C_n/C_p$ ratio that does not 
appreciably deviate from unity at any values of $q$. 
The enhancement of the $C_n/C_p$ ratio in the low-$q$ region, 
observed in Figs.\ \ref{fig:lednicky-cuts}f and \ref{fig:lednicky-cuts-2}f, 
could be interpreted as an indication that for those 
$np$ pairs, protons are on the average emitted earlier 
than neutrons. 

The result of a shorter average emission time for 
protons seems to corroborate our previous findings from $nn$ 
and $pp$ data compared to a two-component 
statistical model \cite{NPA00}.    
However, the calculation of the $C_n/C_p$ ratio with 
the ``two-source model'' parameters used in Ref.\ \cite{NPA00} 
yields a negligible effect (shown in Fig.\ \ref{fig:csorgo}f) as 
compared to the experimental data 
(Figs.\ \ref{fig:lednicky-cuts}f and \ref{fig:lednicky-cuts-2}f). 
While this might indicate that the initial time distribution 
of the ``two-source model'' calculation is too broad and/or 
wrongly shaped, one should also keep in mind that 
there could be alternative explanations.  
The model of \cite{NPA00} neglects 
any residual interaction between the emitted 
particles and the reaction zone. 
In particular, the Coulomb interaction between the proton and the 
emitting source - which could increase the average distance 
between the two particles if the proton is emitted first - 
has not been taken into account \cite{NPA00}, 
although, according to Ref.\ \cite{Lednicky}, 
the influence of the Coulomb field of the residual nucleus 
on the proton is expected to be minimal. 
Alternative explanations might be found in 
reaction mechanisms that are neglected in our calculations, 
such as delayed feeding from the decay of 
excited primary fragments. 
It thus appears as if a refined theoretical treatment 
would be welcome to confirm or not the interpretation 
presented here. 

In conclusion, the experimental results from differently gated 
correlation functions are in qualitative agreement with a 
classical argumentation and calculations using the Koonin-Pratt 
formalism \cite{KoonPrat}, and support the interpretation that 
in high-parallel-velocity and high-total-momentum selected 
events, which enhance projectile-like and/or intermediate 
velocity sources, the proton is on average emitted earlier 
than the neutron.
Although alternative explanations of the enhancement in the 
$C_n/C_p$ ratio cannot be ruled out at this time, 
the experimental evidence of such an 
enhancement, presented here for the first time, 
is by itself intriguing and deserves further experimental 
and theoretical investigation. 

The authors would like to thank Prof.\ S.\ Mrowczynski for 
valuable discussion. The support from the Swedish Natural 
Science Research Council is appreciated. 



\end{document}